\documentclass{jpsj-suppl}
\usepackage{txfonts} 

\title{MCNP6 Simulation of Reactions
of Interest to FRIB, Medical, and Space Applications}

\author{Stepan G. \textsc{Mashnik}}

\inst{Los Alamos National Laboratory, Los Alamos, NM 87545, USA}

\email{mashnik@lanl.gov}

\recdate{July 10, 2014}

\abst{The latest, production, version of the 
Los Alamos Monte Carlo N-Particle transport code MCNP6
has been used to simulate a variety of 
particle-nucleus and nucleus-nucleus 
reactions
of academic and applied interest to 
the Facility for Rare Isotope Beams (FRIB),
medical isotope production, space-radiation shielding, 
cosmic-ray propagation, and accelerator applications,
including several reactions induced by
radioactive isotopes, analyzing production of both stable and radioactive
residual nuclei.
Here, we discuss examples of validation and verification 
of MCNP6 compared to recent neutron spectra
measured at the Heavy Ion Medical
Accelerator in Chiba,
Japan; to
spectra of light fragments from
several reactions measured recently at 
GANIL, France; 
INFN Laboratori Nazionali del Sud, Catania, Italy;
COSY of the J\"{u}lich Research Center, Germany;
and to cross sections of products from
several reactions measured lately
at GSI, Darmstadt, Germany;
ITEP, Moscow, Russia;
and, at 
LANSCE, LANL, Los Alamos, USA.
As a rule, MCNP6 provides quite good predictions for most
of the reactions we analyzed so far,
allowing us to conclude that 
it can be used as a reliable and useful simulation tool for 
FRIB, medical, and space applications involving
stable and radioactive isotopes.}

\kword{
MCNP6, cascade-exciton model (CEM),
Los Alamos version of the quark-gluon string model (LAQGSM),
 FRIB, medical isotope production, space-radiation
applications}

\begin{document}
\maketitle

\section{Introduction}

During the past several years, a major effort has been undertaken at the 
Los Alamos National Laboratory (LANL) to develop the transport code 
MCNP6 \cite{MCNP6, MCNP6.1.1},
the latest and most advanced Los Alamos Monte Carlo transport code representing 
a merger of MCNP5 
and MCNPX. 
Extensive 
Validation and Verification (V\&V) of our low energy transport code MCNP5 has 
been performed and published for many different test-problems involving 
interactions of neutrons, photons, and electrons with thick and thin targets, 
therefore V\&V of MCNP6 for such problems is important but not very critical. 
On the other hand, our high-energy transport code, MCNPX, was not tested 
against experimental data so extensively, especially for high-energy processes 
induced by protons, and heavy-ions. More important, MCNP6 uses the latest 
modifications of the cascade-exciton model (CEM) and of the Los Alamos version 
of the quark-gluon string model (LAQGSM) event generators CEM03.03 and 
LAQGSM03.03 \cite{5}; they were not tested extensively in MCNPX. This is why 
it is necessary to V\&V MCNP6 at intermediate and high energies, to test how 
CEM03.03 and LAQGSM03.03 work in MCNP6 and to make sure that the latter properly 
transports energetic particles and nuclei through matter. Such a V\&V work
has been performed recently for many various reactions
and results were documented in several
publications (see, e.g.,  \cite{EPJP2011, MPIVaV13} and references therein). 

A direct ``trigger" for the current V\&V work
presented here was a recent
Report 
on Benchmarking Heavy Ion Transport Codes FLUKA, HETC-HEDS, MARS15, MCNPX, 
and PHITS
\cite{Remec13}, where not so
good results by an old version MCNPX using an old version of LAQGSM 
were shown
and 
even worse results by a version of MARS15 using a version of LAQGSM 
were published: We need to check how the production version of MCNP6 
describes such reactions.   

\section{Results}

We have tested with the production version of MCNP6 \cite{MCNP6.1.1}
almost all
neutron spectra 
 measured at the Heavy Ion
Medical Accelerator in the Chiba (HIMAC) facility of the National Institute
of Radiological Science (NIRS) 
\cite{Heilbronn07, Nakamura-book06}. Most of our results on this work
have been published in \cite{MPIVaV13}. Fig. 1 presents only two examples 
of $n$ spectra, from 400 MeV/A $^{14}$N + $^{12}$C and $^{64}$Cu. Similar
results were obtained ( see \cite{MPIVaV13}) with the production version of MCNP6 for   
other measurements published in \cite{Heilbronn07, Nakamura-book06}.

\begin{figure}[tbh]
\centering
\includegraphics[width=0.5\textwidth]{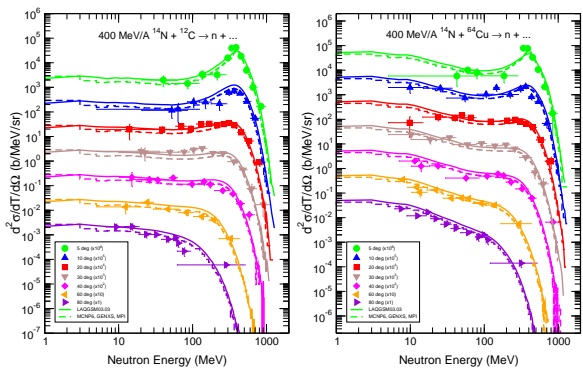}
\caption{
Experimental \cite{Heilbronn07, Nakamura-book06}
neutron spectra from 400 MeV/A $^{14}$N + $^{12}$C and $^{64}$Cu,
compared with calculations by the production version of MCNP6 \cite{MCNP6.1.1}
and by its LAQGSM event generator used as a stand alone code \cite{5},
as indicated.
}
\label{f1}
\end{figure}

\vspace*{-3mm}
Our current good
results obtained with the 
 production version of MCNP6 for   
the measurements published in \cite{Heilbronn07, Nakamura-book06}
make questionable to us the quite poor results by 
a version of LAQGSM used
in a version of MCNPX and even worse results 
published in
Ref. \cite{Remec13}
by MARS15 using a version of LAQGSM. We can assume 
that either some errors were
present in the input files used in the calculations performed in Ref.
\cite{Remec13}, or 
a questionable version of LAQGSM was used
in calculation of the
results published in Ref. \cite{Remec13}.

Fig. 2 shows another example of reactions of interest to FRIB,
namely fragmentation of $^{64}$Ni at 140 MeV/nucleon on $^{9}$Be
target-nuclei measured at MSU for RIA  \cite{Mocko}
compared with results by LAQGSM03.03 as implemented in
MCNP6 \cite{MCNP6, MCNP6.1.1}. Many more examples of such reactions,
proper references, and discussions
can be found in Ref. \cite{LAQGSM03.03}. 
A comparison of our results for such reactions with calculations by several
other models can be found in Ref. \cite{EPJP2011}.
We can see that MCNP6 using
LAQGSM03.03 describes these reactions very well.

Ref. \cite{Varenna2003} presents examples of several reactions 
of interest to FRIB, induced by
radioactive isotopes, measured at GSI and calculated 
by LAQGSM03.03 as implemented in
MCNP6 \cite{MCNP6, MCNP6.1.1}. Those reactions were also describes
well enough by LAQGSM03.03.

Let us note that a recent V\&V of
the production version of MCNP6 \cite{MCNP6.1.1} published in
Ref. \cite{Firenze2013} proved that MCNP6 describes well also 
spectra of light fragments from
several reactions measured recently at 
GANIL
\cite{Dudouet2014};
INFN, Catania \cite{DeNapoli2012}; and
COSY, J\"{u}lich
\cite{Budzanowski2010}.

We have analyzed with MCNP6 or/and with its
event generators
practically all 
cross sections of products from spallation, fission, and fragmentation
reactions induced by protons and nuclei
measured lately
at GSI, Darmstadt, Germany;
ITEP, Moscow, Russia;
and, at 
LANSCE, LANL, Los Alamos, USA. Examples of our results on
these studies and
further references can be found in 
\cite{5, EPJP2011, MPIVaV13, LAQGSM03.03, Firenze2013, p800Th, Titarenko2003}.
Fig. 3 shows only one example on $^{225}$Ac and other medical isotopes 
production
measured and studied recently at LANL \cite{p800Th} using MCNP6.
Calculations with different event-generators of MCNP6 agree well with
measured data, encouraging in light of MCNP6's ability to accurately
predict yields of $^{225}$Ac for medical radiotherapy (see details
in \cite{p800Th}). 

\begin{figure}[tbh]
\centering
\includegraphics[width=0.4\textwidth]{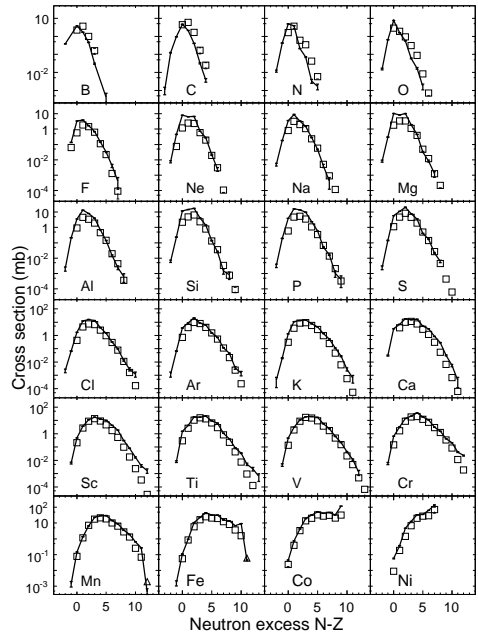}
\caption{Measured cross sections for $^{64}$Ni fragmentation on $^{9}$Be at
140 MeV/nucleon \cite{Mocko} compared with results by a version
of LAQGSM03.03 as implemented in the production version of MCNP6 \cite{MCNP6.1.1}.
}
\label{f1}
\end{figure}

\vspace*{-3mm}

\begin{figure}[tbh]
\centering

\vspace*{-3mm}
\includegraphics[width=0.4\textwidth,angle=-90]{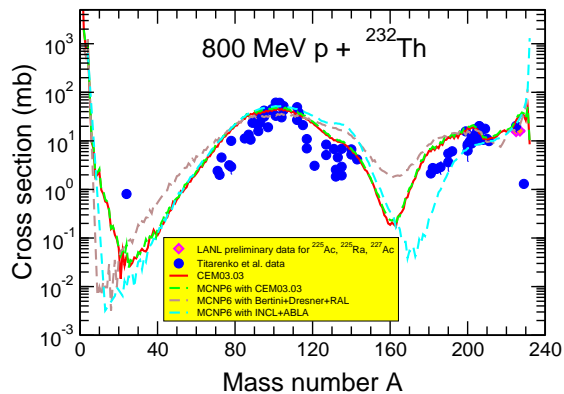}
\caption{
Comparison of the mass distributions of products from 800 MeV p + $^{232}$Th
calculated with MCNP6 using several of its event-generators 
(see details in \cite{p800Th})
with the 
cumulative cross sections measured at ITEP
by Titarenko {\it et al.} \cite{Titarenko2003} and with the recent LANL data
\cite{p800Th}.
}
\label{f1}
\end{figure}

We have tested lately MCNP6 on a number of reactions of interest to
space-radiation shielding and cosmic-ray propagation. One example of results
from our study
is presented in Fig. 4, where we compare the MCNP6 results with the
experimental yields of products from interactions
$\pi^+$ with Fe at the highest
energy measured so far for this reaction \cite{Ferrari2014}, of 120 GeV/c.

\begin{figure}[tbh]
\centering
\includegraphics[width=0.3\textwidth,angle=-90]{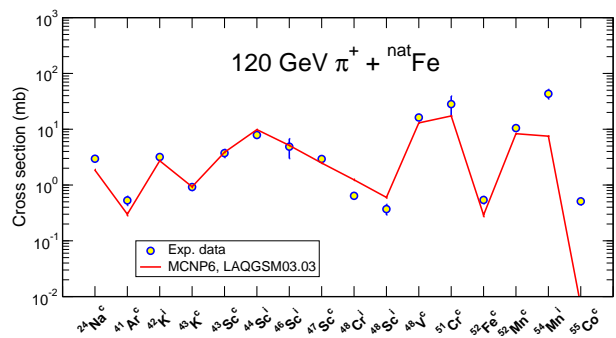}
\vspace*{5mm}
\caption{
Detailed comparison between all cross sections measured in 
Ref. \cite{Ferrari2014} from interaction of 120 GeV/c $\pi^+$
with Fe and those calculated by MCNP6, as indicated.
}
\label{f1}
\end{figure}

\section{Conclusions}

Our study shows that the production version of
MCNP6 provides quite good predictions for most
of the reactions we analyzed so far,
allowing us to conclude that 
it can be used as a reliable and useful simulation tool for 
FRIB, medical, and space applications involving
stable and radioactive isotopes.

This work was supported by the US DOE at LANL under Contract No. DE-AC52-06NA253996.

\end{document}